Templated grain growth in macroporous materials

Florian Bouville[1,3], Etienne Portuguez[1,2], Yunfei Chang[2], Gary L. Messing[2], Adam J. Stevenson[1], Eric Maire[3], Loic Courtois[1,4], Sylvain Deville[1]

[1] Laboratoire de Synthèse et Fonctionnalisation des Céramiques, UMR3080 CNRS/Saint-Gobain, Cavaillon, France

[2] Department of Materials Science and Engineering, Materials Research Institute, Pennsylvania State University, University Park, Pennsylvania 16802, USA

[3] Université de Lyon, INSA-Lyon, MATEIS CNRS UMR5510, Villeurbanne, France

[4] now with: Manchester X-Rays Imaging Facility, The University of Manchester, Oxford Road, Manchester M13 9PL, UK

**Abstract**

We demonstrate a facile method to produce crystallographically textured, macroporous materials using a combination of modified ice templating and templated grain growth (TGG). The process is demonstrated on alumina and the lead-free piezoelectric material Sodium Potassium Niobate (NKN). The method provides macroporous materials with aligned, lamellar ceramic walls which are made up of crystallographically aligned grains. Each method showed that the ceramic walls present a long range order over the entire sample dimensions and have crystallographic texture as a result of the templated grain growth process. We also present a modification of the March-Dollase equation to better characterize the overall texture of materials with textured but slightly misaligned walls. The controlled crystallographic and morphologic orientation at two different length scales demonstrated here can be the basis of multifunctional materials.

**I. Introduction**

Single crystals are used in many applications to take advantage of crystallographic anisotropy that maximizes the physical properties of materials along specific crystal directions. Single crystals are, however, frequently difficult to grow with adequate size, shape, compositional homogeneity, and cost. For this reason, a group of processing strategies has been developed to produce polycrystalline ceramics with crystallographically aligned, or textured, grains.

One particularly promising approach that has been widely applied is Templated Grain Growth[1] (TGG). For TGG, a small population of large template particles (5-50 µm) is dispersed in a fine particle matrix. The templates are aligned during green forming, usually by tape-casting, and epitaxial grain growth from the templates results in microstructures constituted of large and aligned grains. The texture quality of the final material is a function of the template alignment after green forming, the composition of the system, and the sintering conditions.

To date, TGG and texturing methods have produced monolithic ceramics and microstructure composites with improved properties (e.g. mechanical, piezoelectric) relative to random ceramics. Such processing strategies are nevertheless unable to produce some types of composite architectures that combine controlled macroscopic porosity with textured ceramic layers. We propose here a modified ice templating approach to fabricate porous, composite TGG structures.

Ice templating is a well-established processing route for porous materials[2,3] where a temperature gradient across a suspension causes ice crystals of the liquid to nucleate and grow, rejecting the solid particles and forming the template of a porous structure. Subsequently, the ice crystals are removed by sublimation, leaving behind pore spaces separated by densely packed walls of the primary particles from the original suspension. Subsequent processing, like sintering, can be used to control the microstructure of the particle walls imparting strength or other desired physical properties.

Ice templating is also a promising approach for hierarchical assembly of complex structures[4,5] because it can be used to simultaneously control several different size scales in the materials. Particles on the nm-µm scales are consolidated within walls that are µm-mm scale, and the samples obtained have dimensions in the millimeter to centimeter range. This hierarchical aspect makes it an ideal tool to realize composites with various connectivities and, at the same time, control the microstructure of each component. For instance, ice-templating was used to obtain 3-1 piezoelectric composites[6] where the ceramic phase was a randomly oriented polycrystal. Such materials have the electromechanical coupling coefficient of a bulk material combined with adaptive acoustic impedance.

In order to combine the composite architectures made possible by ice templating with oriented ceramics made by TGG, it is necessary to align template particles during the ice templating process. Recently, anisotropic particles such as graphene[7] and alumina platelets[8] were aligned by ice templating, but the alignment has always been obtained only locally; anisotropic particles within each wall were aligned relative to one another, but misorientation between the individual walls meant that the texture was not observed over the entire sample . In order to extend the alignment developed within the individual walls to the entire sample, the walls need to be ordered over the entire sample. Such a lamellar structure can be the basis for 2-2 and 3-2 composites. Some modifications of the process have been developed to obtain lamellar pores in samples of large dimensions[9] but have proved hard to replicate.

In this study we present a method that combines ice templating and TGG. We first demonstrate how to make textured, macroporous alumina as a model system and then apply the process to sodium potassium niobate, NKN, a promising lead-free piezoelectric composition. We also show that the ceramic walls produced during ice templating can be aligned based on the application of a second thermal gradient during the freezing to produce 2-2 type composites with ordered lamellar macropores between textured NKN layers. Because of the complex interconnections and morphologies of the textured and ice templated materials, standard tools like XRD rocking curve analysis or Lotgering factor are insufficient to characterize the structure. Therefore, we propose a different strategy that couples X-rays diffraction (XRD) and morphological descriptors with a modification of the March-Dollase equation.

**II. Experimental procedure**

For the alumina samples, commercially available alumina platelets (RonaFlair White Sapphire; produced by Antaria ltd, sold by Merck KGaA) with 8 µm diameter and 500 nm thickness were used as template particles and 100 nm diameter isotropic alumina (TM-DAR, Taimicron) as matrix particles. Nanoparticles of silica (Nexsil 20K, diameter 20 nm, provide by Nyacol) and calcium carbonate (Sigma Aldrich) were added as liquid phase precursors in 75:25 molar ratio of $SiO_2$:CaO. The matrix powder, liquid phase precursors, and a binder (PEG20M, 4 wt. % of the powder mass, Sigma Aldrich) were mixed with distilled water and ball-milled for 20h. The template particles were added only three hours before the end of the milling step to avoid any excessive breakage of the platelets.

For conventional ice templating, the slurry contained 20 vol.% of solid which comprised 9.8 vol% of templates, 88.5 vol.% of matrix powder, and 1.7 vol.% of liquid phase $SiO_2$+CaO. The slurry was then poured into a cylindrical Teflon mold and placed in contact with a copper cold plate. Temperature was controlled by circulating refrigerated silicone oil through the copper plate, and the freezing rate was -1°C/min. Once the sample was completely frozen, it was removed from the mold and freeze dried for at least 48 h in a commercial freeze-dryer (Free Zone 2.5 Plus, Labconco, Kansas City, Missouri, USA) to ensure a complete removal of the ice crystals.

In order to align the ceramic walls during ice-templating, the slurry was frozen as it flowed over the freezing surface. A rectangular mold (2 x 10 x 3 $cm^3$) made of silicon rubber was placed on an inclined copper plate. The slurry was then introduced at the top end of the mold and allowed to flow down the inclined plane of the copper plate as it froze. Slurry that reached the bottom of the mold was recirculated using a peristaltic pump. The flow was maintained at 70 cm$^3$/min until freezing was complete. A cooling rate of -1°C/min was used for all the samples. Once a first layer (of around 5 mm in thickness) of aligned crystals was frozen, the slurry recirculation was turned off, the copper plate was leveled, and the mold was filled with the remaining slurry. A thermal camera (ThermaCAM E45, FLIR, Croissy-Beaubourg, France) was used to measure the temperature of the flowing suspension during the solidification of the first layer. This camera has a sensitivity of 0.1°C and can measure temperature down to -10°C. Once freezing was completed, the samples were freeze-dried under conditions identical to those used for conventional ice templated alumina.

The sintering cycle of alumina samples frozen at -1°C/min was chosen in accordance with the study by Pavlacka et al.[10] with a sintering step at 1550°C for 90 minutes. For SEM observations, the samples were fractured and thermally etched at 1450°C for 20 minutes to better reveal the grain boundaries.

For NKN samples, anisotropic sodium niobate (NN$_T$) platelets 15 µm in diameter and 1 µm thick were synthesized for use as template particles. Sodium potassium niobate (NKN) particles 0.5 µm in diameter were produced to be used as the matrix particles. The NN$_T$ particles were prepared by Topochemical Microcrystal Conversion[11,12] of bismuth niobate seeds according to the procedure detailed by Chang et al[13]. The NKN particles were made by a conventional mixed oxide synthesis [13]. 1 mol % of copper oxide (Sigma Aldrich) was added as liquid phase precursor.

The slurry preparation conditions and freezing conditions for NKN were identical to that of alumina. The sintering cycle of the NKN sample was chosen in accordance with the study of Chang et al.[13], with a sintering step at 1115°C for 4h and an annealing step at 1125°C for 3h.

SEM pictures were taken with a Supra 55 microscope (Zeiss) on gold coated samples. Samples were scanned on a Phoenix tomograph with a 3.4 µm$^3$ voxel size on a disk-shaped sample of 3 mm diameter and 3 mm in high. The directionality measurement was made with the plug-in named "directionality" of the free software Fiji[14]. XRD measurements were performed on a D5000 device (Brüker).

**III. Results and discussions**

**(1) Platelet Alignment in a Particle Matrix by Ice Templating**

There are several examples where the growth of ice crystals effectively organizes anisotropic particles[5,7,8], but for TGG a minor fraction of platelets must be aligned within a powder matrix. Fig. 1a shows that the platelets in the green body are generally aligned in the same direction as the ceramic wall, although some degree of misorientation is observed. A recent study[15] has shown that compaction between the growing crystals aligns the platelets, similar to the alignment of platelets during unidirectional pressing. The most important differences between the two situations are the magnitude of the compaction force and the fact that ice templating takes place in a colloidal suspension. The key parameters to act on platelet orientation are interparticle forces, steric impingement, and, in some special cases, preferential surface interactions between the platelets/ice crystals.

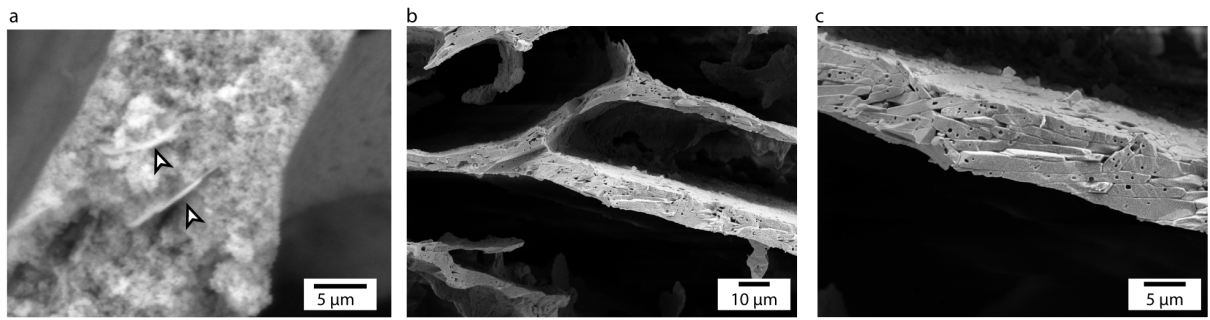

Fig. 1: (a) SEM observations of the platelets (black and white arrows) alignment in the green body. (b) Close view of wall's organization after sintering, showing the correlation between wall and grain orientation. (c) SEM picture showing the high grain alignment.

The microstructure after sintering is shown in Fig. 1b and Fig. 1c. These samples have a lamellar structure where 10-20 µm thick lamellar pores are separated by 5-10 µm thick ceramic layers. In Fig. 1b, large tabular grains are easily observed within the ceramic walls, and the orientation of such grains is generally parallel to the border of the walls. The grains keep their preferred orientation even when the wall itself is divided or wavy. Fig. 1c shows a higher magnification image of the ceramic walls, and most of the grains in the wall are aligned within 15° of the wall orientation. This measure was evaluated by analyzing the grain orientation on SEM pictures[16] within the ceramic walls on thermally etched microstructures.

**(2) Control of the lateral wall orientation by freezing under flow**

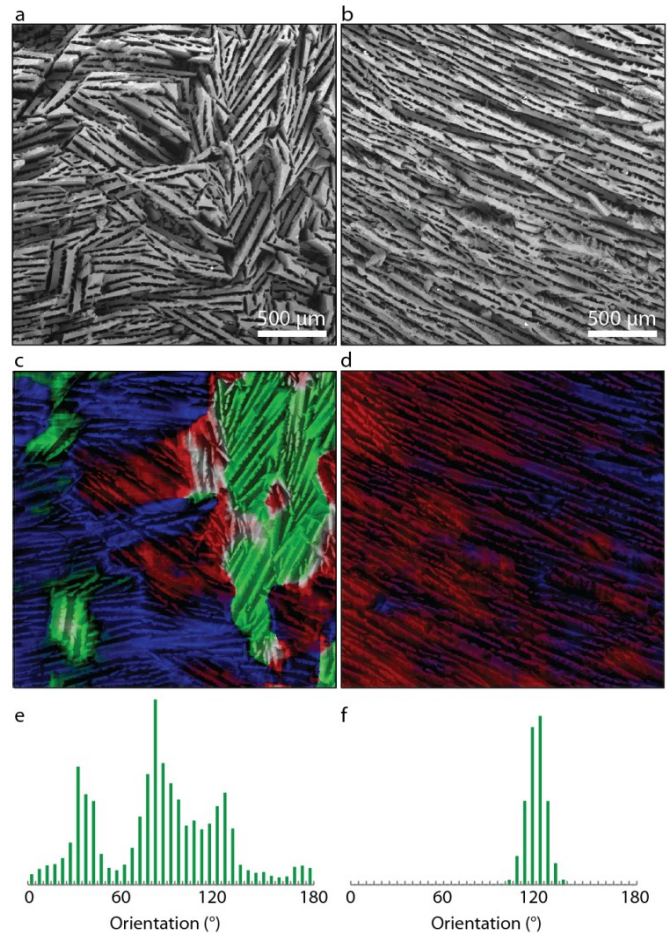

Fig. 2: Comparison of randomly oriented ice-templated microstructure (a) and (b) oriented sintered sample made by freezing under flow. Domains with long range ordered walls are represented with the same color in (c) and (d) for each sample, see reference[16] for details on the method. (e) and (f) represent the angular orientation histogram of pixel gradient in the image.

Fig. 2a shows a microstructure typical of ice templated, sintered ceramics. The sample microstructure is divided into individual domains where the walls present a short range order, but the orientation distribution of the walls over the entire sample is random (Fig. 2c and e). No long range order is present. While the images in Fig. 1 show that TGG is possible within individual walls by ice templating, we must also align the walls to benefit from texture over the whole sample.

Controlling the lateral orientation of the ceramic walls is an important challenge in ice templating and is critical to realizing the improved properties we seek from the textured ceramic walls. Previously, the alignment of ice templated walls has been achieved by combining a patterned ice nucleation surface with gentle vibration during freezing[9,17]. Because this process is inherently linked to a specific freezing setup and has proven difficult to replicate, we developed a different and hopefully more robust alignment technique.

Ice crystal orientation is intimately related to the direction and magnitude of the applied temperature gradient[18]. Adding a second gradient could thus orient the crystal growth in a direction imposed by the combination of the two gradients. Based on this idea, we developed a simple process, referred to as "freezing under flow."

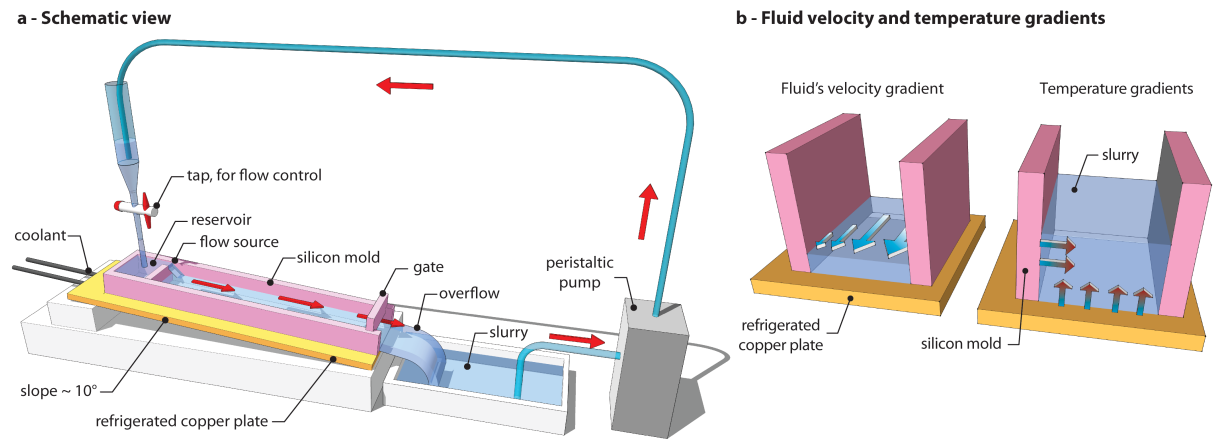

Fig. 3: Freezing under flow: setup and principles. (a) Schematic representation of the freezing under flow setup (b) Schematic representation depicting the fluid velocity and temperature gradients.

The principles of freezing under flow are schematically depicted in figure 3a. A temperature gradient is naturally generated as the slurry flows along a refrigerated copper plate. The suspension overflow at the open end of the mold is re-circulated using a peristaltic pump and a reservoir. The flow is carefully controlled to avoid the introduction of bubbles during the process. Because the flow source is located at one side of the mold, a parabolic velocity gradient occurs during flow (Fig. 3b), and the slurry cools more rapidly along one side of the mold (where it is in contact with the refrigerated plate for a longer time). Thermograms (Fig 4a) taken at the beginning of the freezing process showed that the temperature difference between the side and the center can be up to 20°C with a maximum lateral temperature gradient approaching 50°C/cm. Because of the location of the flow source, the temperature profile is asymmetric (Fig. 4d).

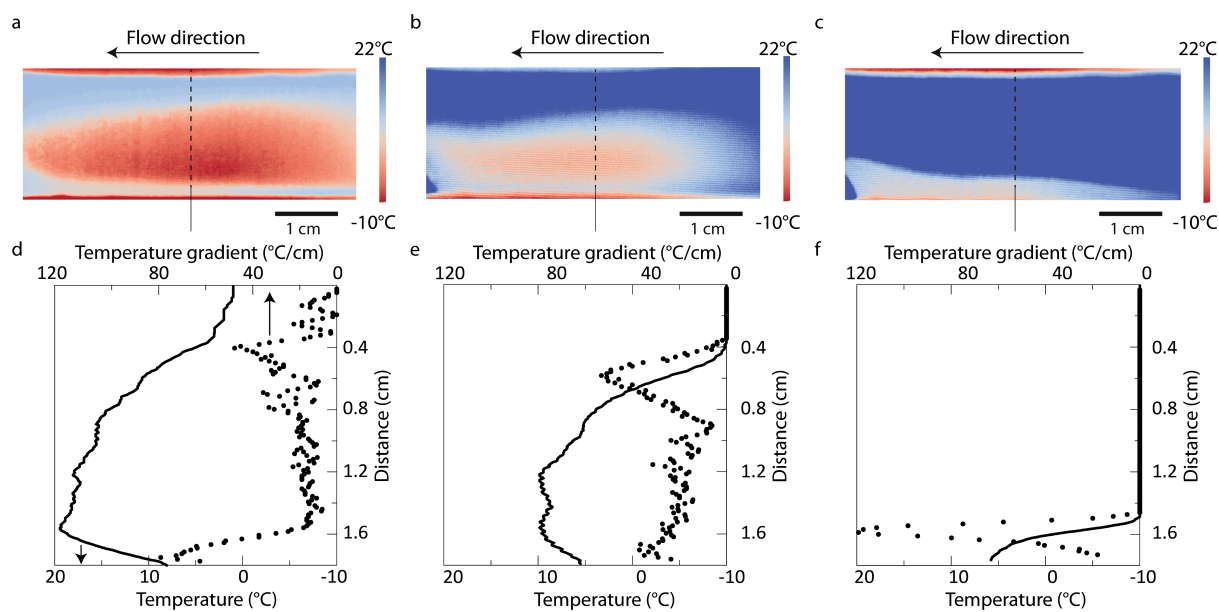

Fig. **4**: Thermal measurement during freezing under flow. The images represent the temperature of the flowing suspension at three different times: before the solidification starts (a), when nearly half of the layer is frozen (b), and before the end (c). The plots (d), (e), and (f) represent the temperature profiles in the middle of the sample and their derivative as a function of the distance for each selected time. The flow source is located in the lower right corner of each thermogram.

During ice templating, the local temperature gradient dictates the solidification front velocity and consequently the ice crystal size[18], so the microstructure of ice templated samples may change if the temperature gradient changes during the process. Fig. 4e shows the temperature profile and lateral gradient after 5 minutes of freezing under flow, and the maximum lateral temperature gradient remains around 50 °C/cm (Fig. 4e). The maximum lateral temperature gradient increases significantly in the last solidification steps, where it can reach 120°C/cm (Fig. 4f) because there is no longer any fluid flow, which helps maintain the constant temperature gradient throughout most of the process. Using the freezing under flow method, the microstructure is similar from one side to the other in the material due to the fact that the process is self-regulating and maintains the same lateral temperature gradient. Domains can still be found in the microstructure (Fig. 2d) but their orientation mismatch is small compared to samples frozen under the typical static conditions (Fig.2e and Fig.2f). Thus, a significant benefit of the freezing under flow process is that the maximum lateral temperature gradient is constant throughout most of the process. Once the entire width of the mold is frozen, the gate at the open end of the silicon mold is closed, the mold is filled with the remaining suspensions, and freezing continues until completion.

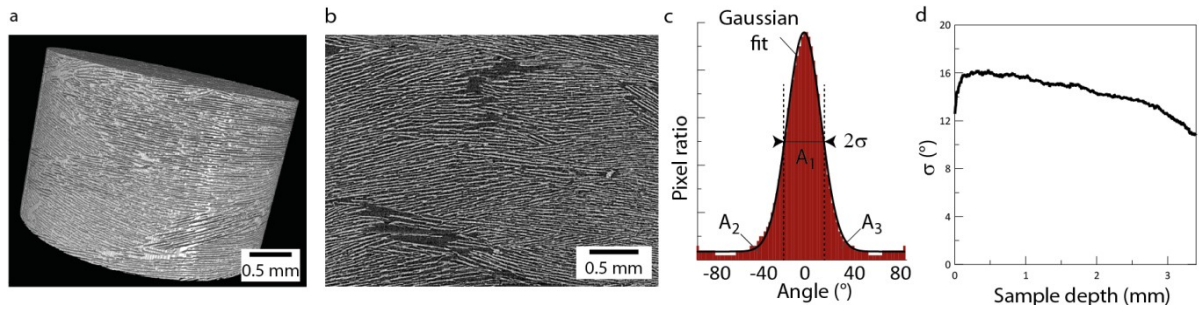

Fig. **5**: Three-dimensional microstructure and morphological characterization of a sample made by freezing under flow. (a) Tomographic reconstruction of the sample. (b) Typical section of the reconstruction on which analysis is performed. (c) Histogram of the pixel fraction presenting a preferred orientation versus the considered angle of the image (b). A Gaussian fit is applied to this distribution and the lateral dispersion σ, the amount of textured pixel and the fit accuracy are extracted. (d) Lateral dispersion as a function of the position in the sample.

In order to characterize the alignment between the ceramic walls, we imaged the samples using X-ray tomography (Fig. 5a), computed the wall orientations[16], and fit the data with a Gaussian distribution to obtain the wall orientation distribution function (Fig 5b and c). We also evaluated the wall orientation distribution function as a function of depth in the sample. The standard deviation of the orientation distribution function shifts from 16° to 10° (Fig. 5d) in the solidification direction. Such dispersion could arise from the sample preparation or microstructure evolution during the freezing. We note that the changing microstructure observed here is in the vertical direction, and is thus not caused by the lateral temperature gradient imposed by the freezing under flow process. In all measurements, more than 80% of the sample volume is textured. Freezing under flow is thus able to induce a long range order of ice crystals over a large sample by creating a second self-regulating gradient during the solidification.

**(3) Texture measurement in composite, layered materials**

The final texture of the composite is a result of the superimposition of two distinct alignments present at two length-scales: (1) the grain alignment in the wall and (2) the wall alignment in the composite. The orientation of grains in bulk textured materials can be measured by several XRD methods, like pole figure or rocking curve[19], but these methods provide information on the first microns below the sample surface due to the limited depth of penetration of X-rays, which in our case corresponds only to the first wall. Thus, XRD based methods are able to measure the grain alignment within the wall, but they cannot characterize the alignment of the ceramic walls throughout the composite. Any functional property of a textured material is directly linked to the dispersion of crystallographic orientation of the whole sample, not just the texture measured in the first few layers. Thus, XRD measurement only provides partial information on the texture of the macroporous sample synthetized here.

One method of characterizing the orientation distribution in materials made by templated grain growth is performing rocking curve measurements and fitting the data to the March-Dollase (MD) distribution[10,19] (1).

$$MD(f, r, \omega) = f \left( r^2 cos^2\omega + \frac{sin^2\omega}{r} \right)^{-3/2} + (1 - f) \qquad (1)$$

The porous material made here can be considered as a combination of layers, each one of those slightly misoriented with respect to the surface (2), as depicted in Fig. Fig. 6a. Equation 1 can be extended to account for the misalignment of an individual ceramic wall by introducing an angle, $\alpha_n$, which characterizes the angle between the ceramic walls and the reference frame of the sample/measurement:

$$W = f \left( r^2 cos^2(\omega - \alpha_n) + \frac{sin^2(\omega - \alpha_n)}{r} \right)^{-3/2} + (1 - f) \qquad (2)$$

Because α represents the shift between the reference frame and the wall orientation, it follows the Gaussian distribution of wall orientations determined by tomography (Figure 4). By adding a large number of MD functions with f and r parameters determined for a single wall by XRD, but shifted by an angle, following the Gaussian distribution determined in Figure 4, one can model the behavior of a textured composite (3).

$$Composite = \sum_1^{Nwall} f \left( r^2 cos^2(\omega - \alpha_n) + \frac{sin^2(\omega - \alpha_n)}{r} \right)^{-3/2} + (1 - f) \qquad (3)$$

$\alpha_n \sim N(0, \sigma)$, where N is a Gaussian distribution of mean 0 and standard deviation σ.

As we know the MD functions and the Gaussian distribution of the walls, we can calculate the FWHM of the composite distribution but also model its evolution with the Gaussian and MD parameters. The FWHM of our alumina composite is 36° (see Fig. 6b), which is relatively high compared to the measurements made on tape cast materials and even compared to the theoretical grain orientation (see Fig. 6b). This value is greater than both the FWHM of Gaussian distribution and MD function, 28° and 5.7° respectively.

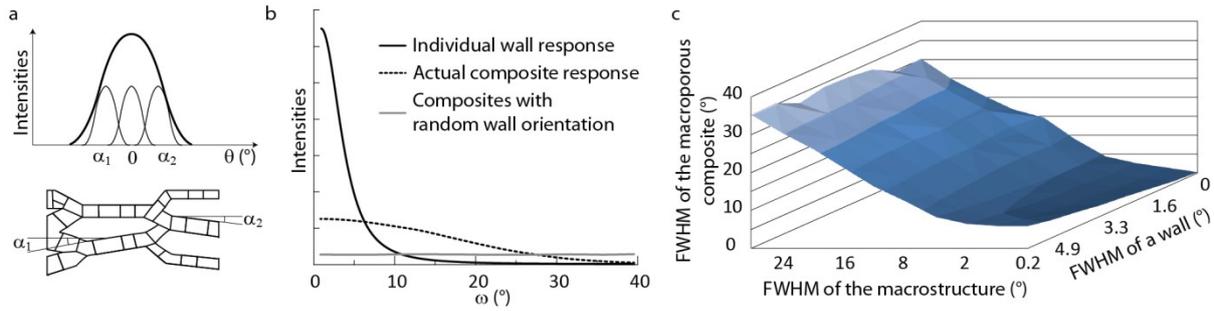

Fig. 6: (a) Representation of the cubic grain textured wall slightly misaligned and the contribution of those walls in the material global orientation distribution. (b) Example of composites texture response for different conditions. (c) Evolution of the FWHM of the composite with respect to the FWHM of the constituent.

To better understand the interplay between the wall alignment and grain alignment within the walls, we calculated the evolution of the composite FWHM versus the two parameters, the FWHM of the macrostructure of the walls and the FWHM of the grains within one individual wall (Fig. 6c).

As expected, reducing the FWHM of both distributions tends to decrease the FWHM of the composite. From our experimental data, it is clear that better aligning the individual ceramic walls during ice templating, rather than improving the texture within the walls, is more important for obtaining high texture quality composites in the future. The wall orientation distribution function could be reduced by optimizing the freezing under flow process by increasing the magnitude of the lateral temperature gradient (thus an increase in the flow velocity). Indeed, the texture quality within the walls could theoretically be improved by improving template alignment within the walls, but it would be a relatively small increase given that the level of misorientation between the walls themselves is much greater (the smallest FWHM reported to date with alumina is 4.6°[10]).

**(4) Textured piezoelectric composite**

Given the results presented above for alumina, it is clear that ice-templating can texture grains in ceramic walls of a macroporous body and align such walls in two directions in a single processing step. The same process has been applied to a NKN powder in order to demonstrate that textured composites, like those required in some sensor and transducer applications, can be made in a single processing step. The microstructure of the porous sample is presented in Fig. 7a and Fig. 7b. The microstructure obtained (Fig. 7a) is similar to the alumina sample because the ice templating process is not strongly influenced by composition (physical characteristics of the particles, platelets, and the suspension are much more important). The large grains present in the sintered wall (Fig. 7b) are cube shaped, due to the weakly orthorhombic crystallographic structure of NKN. As was the case in alumina, the ice crystal growth effectively aligned the $NN_T$. Dense and textured walls are obtained after sintering.

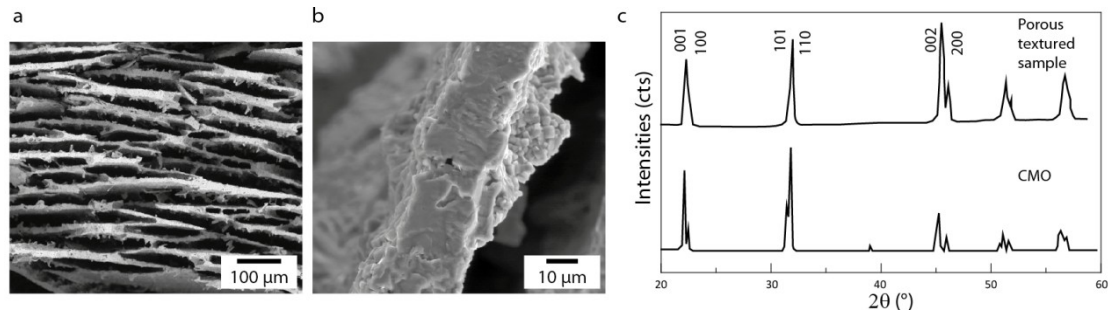

Fig. 7: (a) Sintered Microstructure of sample frozen under flow, showing the alignment of ceramic wall. (b) SEM picture of a wall after sintering showing the large grains. (c) Comparison of XRD pattern obtained with textured composite and non-textured sample (CMO stands for Conventional Mixed Oxide, reproduced from [13]).

The presence of preferred grain orientation has been investigated by XRD measurements. Two XRD patterns are shown in Fig. 7c: the bottom one represents a randomly-oriented sample obtained by conventional mixed oxide NKN powder and the top XRD pattern the porous textured composite. A clear texture of the <00l> plane is present in the sample made by ice-templating. However the texture is not as good as in samples made by tape-casting[13] and the Lotgering factor of the sample is only 21%.

## IV. Conclusions

We demonstrated that ice-templating combined with templated grain growth is an effective method to produce textured, macroporous materials. The growth of ice crystals was used to align platelets that are used afterward as seeds for templated grain growth. To retain the benefits arising from this local texture, a 'freezing under flow' method has been developed to control the long range order of the lamellar ice crystals and thus ensure both short and long range order of the grain orientation. This novel method is robust and independent of the material used. The original dispersion of orientation of the ice crystals compromises the optimization of the functional properties and needs to be reduced, as suggested by the model developed here. Such reduction could be achieved by modifying the freezing conditions, which requires a more comprehensive study of the process parameters. The principles demonstrated here could be the basis for hierarchical, multifunctional composite materials, including structural and piezoelectric ceramics.

**Acknowledgements**

We acknowledge the financial support of the ANRT (Association Nationale Recherche Technologie) and Saint-Gobain through a CIFRE fellowship, convention #808/2010.

**References**

(1)     Messing, G. L.; Trolier-McKinstry, S.; Sabolsky, E. M.; Duran, C.; Kwon, S.; Brahmaroutu, B.; Park, P.; Yilmaz, H.; Rehrig, P. W.; Eitel, K. B.; Suvaci, E.; Seabaugh, M.; Oh, K. S. Templated Grain Growth of Textured Piezoelectric Ceramics. *Crit. Rev. Solid State Mater. Sci.* **2004**, *29*, 45–96.

(2)     Deville, S.; Saiz, E.; Nalla, R. K.; Tomsia, A. P. Freezing as a Path to Build Complex Composites. *Science* **2006**, *311*, 515 –518.

(3)     Munch, E.; Launey, M. E.; Alsem, D. H.; Saiz, E.; Tomsia, A P.; Ritchie, R. O. Tough, Bio-Inspired Hybrid Materials. *Science* **2008**, *322*, 1516–1520.

(4)     Deville, S. Ice Templating, Freeze Casting: Beyond Materials Processing. *J. Mater. Res.* **2013**, *28*, 1–18.

(5)     Bouville, F.; Maire, E.; Meille, S.; Van de Moortèle, B.; Stevenson, A. J.; Deville, S. Strong, Tough and Stiff Bioinspired Ceramics from Brittle Constituents. *Nat. Mater.* **2014**, *advance online publication*.

(6)     Lee, S.-H.; Jun, S.-H.; Kim, H.-E.; Koh, Y.-H. Piezoelectric Properties of PZT-Based Ceramic with Highly Aligned Pores. *J. Am. Ceram. Soc.* **2008**, *91*, 1912–1915.

(7)     Qiu, L.; Liu, J. Z.; Chang, S. L. Y.; Wu, Y.; Li, D. Biomimetic Superelastic Graphene-Based Cellular Monoliths. *Nat. Commun.* **2012**, *3*, 1241.

(8)     Hunger, P. M.; Donius, A. E.; Wegst, U. G. K. Platelets Self-Assemble into Porous Nacre during Freeze Casting. *J. Mech. Behav. Biomed. Mater.* **2013**, *19*, 87–93.

(9)     Munch, E.; Saiz, E.; Tomsia, A. P.; Deville, S. Architectural Control of Freeze-Cast Ceramics Through Additives and Templating. *J. Am. Ceram. Soc.* **2009**, *92*, 1534–1539.

(10)    Pavlacka, R. J.; Messing, G. L. Processing and Mechanical Response of Highly Textured Al2O3. *J. Eur. Ceram. Soc.* **2010**, *30*, 2917–2925.

(11)    Poterala, S. F.; Chang, Y.; Clark, T.; Meyer, R. J.; Messing, G. L. Mechanistic Interpretation of the Aurivillius to Perovskite Topochemical Microcrystal Conversion Process. *Chem. Mater.* **2010**, *22*, 2061–2068.

(12)    Saito, Y.; Takao, H.; Tani, T.; Nonoyama, T.; Takatori, K.; Homma, T.; Nagaya, T.; Nakamura, M. Lead-Free Piezoceramics. *Nature* **2004**, *432*, 84–87.

(13)    Chang, Y.; Poterala, S. F.; Yang, Z.; Trolier-McKinstry, S.; Messing, G. L. Microstructure Development and Piezoelectric Properties of Highly Textured CuO-Doped KNN by Templated Grain Growth. *J. Mater. Res.* **2010**, *25*, 687–694.

(14) Schindelin, J.; Arganda-Carreras, I.; Frise, E.; Kaynig, V.; Longair, M.; Pietzsch, T.; Preibisch, S.; Rueden, C.; Saalfeld, S.; Schmid, B.; Tinevez, J.-Y.; White, D. J.; Hartenstein, V.; Eliceiri, K.; Tomancak, P.; Cardona, A. Fiji: An Open-Source Platform for Biological-Image Analysis. *Nat. Methods* **2012**, *9*, 676–682.

(15) Bouville, F.; Maire, E.; Deville, S. Self-Assembly of Faceted Particles Triggered by a Moving Ice Front. *Langmuir* **2014**.

(16) Jeulin, D.; Moreaud, M. Segmentation of 2D and 3D Textures from Estimates of the Local Orientation. *Image Anal. Stereol.* **2008**, *27*, 183–192.

(17) Launey, M. E.; Munch, E.; Alsem, D. H.; Barth, H. B.; Saiz, E.; Tomsia, A. P.; Ritchie, R. O. Designing Highly Toughened Hybrid Composites through Nature-Inspired Hierarchical Complexity. *Acta Mater.* **2009**, *57*, 2919–2932.

(18) Waschkies, T.; Oberacker, R.; Hoffmann, M. J. Investigation of Structure Formation during Freeze-Casting from Very Slow to Very Fast Solidification Velocities. *Acta Mater.* **2011**, *59*, 5135–5145.

(19) Brosnan, K. H.; Messing, G. L.; Meyer, R. J.; Vaudin, M. D. Texture Measurements in <001> Fiber-Oriented PMN-PT. *J. Am. Ceram. Soc.* **2006**, *89*, 1965–1971.